\documentclass[twocolumn,superscriptaddress,showpacs]{revtex4}
\usepackage{graphicx,amsmath,subfigure}
\usepackage{epstopdf}

\newcommand{\ie}{\emph{i.e.}}

\newcommand{\ER}{Erd\H{o}s-R\'{e}nyi}

\begin{document}

\title{Local cluster aggregation models of explosive percolation}
\author{Raissa M. D'Souza}
\affiliation{University of California, Davis, CA 95616, USA}
\affiliation{Santa Fe Institute, 1399 Hyde Park Road, Santa Fe, New Mexico 87501, USA}
\email{raissa@cse.ucdavis.edu}
\author{Michael Mitzenmacher}
\affiliation{School of Engineering and Applied Sciences, Harvard University, Cambridge, Massachusetts, USA}
\email{michaelm@eecs.harvard.edu}

\begin{abstract}
We introduce perhaps the simplest models of graph evolution with
choice that demonstrate discontinuous percolation transitions and 
can be analyzed via mathematical evolution equations.
These models are {\em local}, in the sense that at each step of the process one edge 
is selected from a small set of potential edges sharing 
common vertices and added to the graph.  
We show that the evolution can be accurately described by a system of differential equations and that such models exhibit the discontinuous emergence of the giant
component.  Yet, they also obey scaling
behaviors characteristic of continuous transitions, with
scaling exponents that differ from the classic Erd\H{o}s-R\'{e}nyi model.
\end{abstract}


\maketitle

The percolation phase transition on both lattices and networks is a subject of intense study, as it provides a model for the onset of large-scale connectivity in random media, such as resistor networks, porous rocks, forest fires, and even social networks~\cite{StaufferPercBook,Solomon2000}. It was recently shown
via numerical simulation of graph evolution obeying an Achlioptas process
that percolation transitions can be discontinuous~\cite{EPScience}.
Starting from a graph of isolated nodes, at each
step of the evolution, two potential edges are chosen uniformly at
random, and using some pre-set criteria one edge is added to the graph
and the other discarded. If 
the edge which minimizes the product or sum of the size of the two components that
would be merged is chosen, then one can show the percolation
transition is discontinuous. Specifically, the size of the
largest component goes from sub-linear in system size $n$ to a large
fraction (bounded away from 0) of the entire network as 
a sub-linear number ($n^{l}$, with $l<1$) of edges are added to the graph.  
Although several recent papers have explored the intuition for the 
mechanisms behind this behavior, such as identifying that the
evolution in the subcritical regime must keep larger components
of similar size~\cite{powderkeg09,Moreira,ChoClusterAgg,Makse2009,Manna},
there are not yet mathematical evolution equations describing the
process.

In contrast, more restricted Achlioptas processes evolving under ``bounded size rules'' can be described mathematically. Such rules are constrained
so that all components of size greater than some cutoff are
treated equivalently. In~\cite{SpencWorm} it was rigorously shown that
graph evolution under bounded size rules
can be accurately described in terms of differential equations.
It is not known, however, whether the restriction to bounded size
rules leads to continuous or discontinuous percolation transitions.

Here we introduce and analyze graph evolution models with choice that are both more physically motivated and simpler mathematically. In contrast to those in~\cite{EPScience}, our models are {\it local} in the sense that the  
choice is constrained to involve edges that share one vertex in common. 
Thus, the candidate edges span up to three components (rather than four as in~\cite{EPScience}). 
We develop a system of differential equations describing the evolution of the components under the associated bounded size rules, which for the simplest model show that the system must reach a critical point.  
We implement our equations numerically and find they
accurately predict the location of the percolation
transition for the case of unbounded rules, and demonstrate
via simulation that the transition for unbounded rules is discontinuous.

\begin{figure}[t]
  \begin{center}
\includegraphics[width=.45\textwidth]{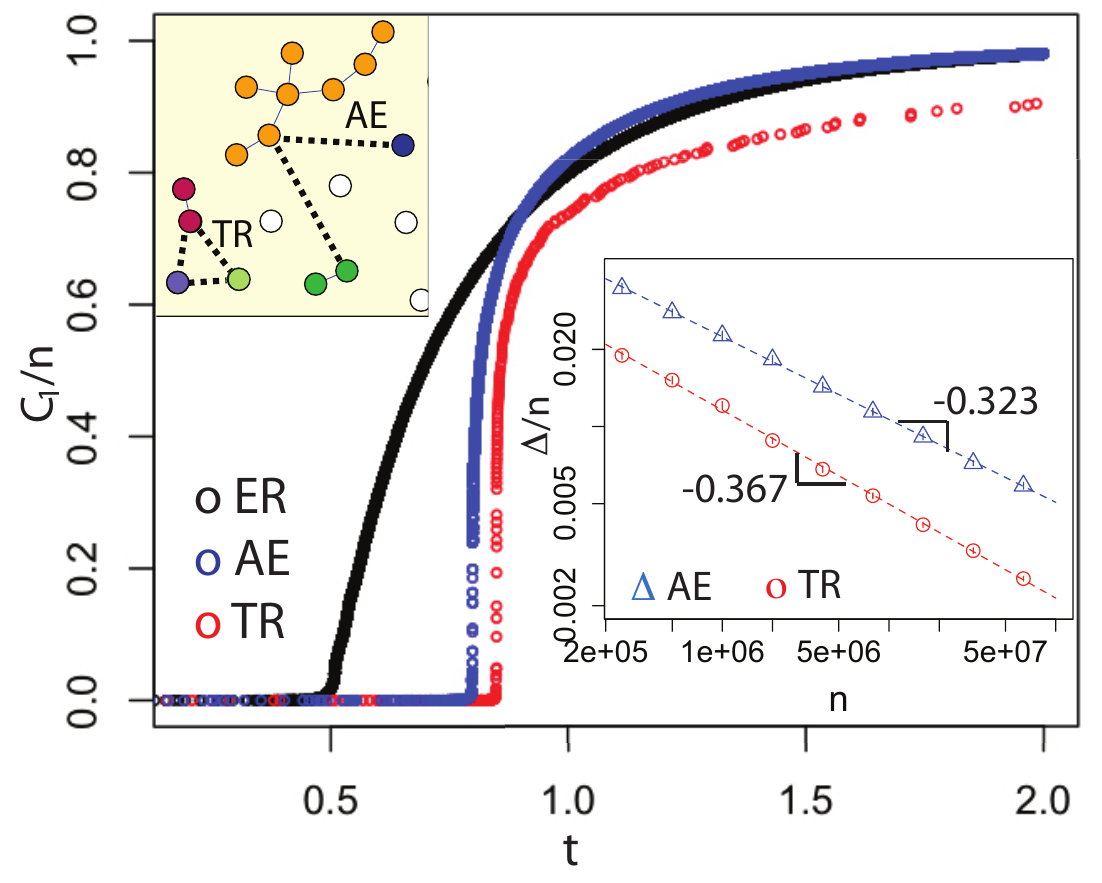}
  \end{center}
  \vspace{-0.2in}
  \caption{Typical evolution of ${C_1}/n$ for 
  \ER\ (ER), Adjacent Edge (AE), and Triangle Rule (TR), for $n=10^6$. 
 (Top inset: Example of three candidate edges for TR, and two candidate edges for AE.)
 (Bottom Inset: $\Delta_n(1/2,0.2)/n$ vs $n$ for AE and $\Delta_n(1/2,0.4)/n$ vs $n$ for TR. Each data point is the average over 50 iid realizations, with error bars smaller than symbols. Dashed lines are $\Delta/n = 1.95 n^{-0.323}$  for AE and $\Delta/n = 1.84 n^{-0.367}$ for TR.)}
\label{fig:evol}
\end{figure}

We explicitly analyze two distinct local processes, although our
approach can naturally be used to examine other similar processes.  We 
call the simplest the adjacent edge (AE) rule: at each step, a first vertex
is chosen uniformly at random, and it must connect to one of two distinct additional vertices also chosen uniformly at random (thus both candidate edges are adjacent). 
Intuitively, the first vertex is forced to connect to one of two random choices.
Here we choose 
the edge that connects it to the additional vertex in the smaller component, except
possibly in the asymptotically negligible case where the first vertex
is in the same component as one of the other two (discussed in detail below). 
Typical evolution of the largest component of the graph, denoted $C_1$,  
is shown in Fig.~\ref{fig:evol}. 
In the bounded size rule version, the same rule above is applied
unless both components for the two additional vertices have size
larger than some bound $K$.  In that case, we simply connect to the first of the
two additional vertices.  

We follow the approach of Spencer and Wormald~\cite{SpencWorm}.
We start with an empty graph $G$ of $n$ vertices.
Let $x_i(G)$ be the fraction of vertices  in components of size $i$:
\begin{equation}
x_i(G) = \frac{1}{n} |\{v : c(v) = i\}|,
\end{equation}
where $c(v)$ is the size
of the component containing $v$. 
Note that $x_i(G) = i n_i(G)$, where $n_i(G)$ is the component density (number of components of size $i$ divided by the system size $n$) often used in cluster aggregation models in the physics literature~\cite{BenNaimPRE05, ChoClusterAgg}.
For bounded size variations we will be interested in $x_i(G)$ for $i \leq K$.

We provide a mean-field analysis over all graph evolutions.  
Hence $x_i$ becomes a function of time, $x_i(t)$, 
where we scale so that a unit of time $1$
corresponds to $n$ edges; we suppress the dependence on $t$ where the
meaning is clear.  We also use $s_j = 1 - \sum_{k < j} x_k$; that is,
$s_j$ is the weight of the tail of the distribution starting from $j$.
The probability the first vertex is in a component of size $i$
is $x_i$.  The probability that the smaller of the two additional components 
has size $j$ is $s_j^2 - s_{j+1}^2 = 2x_j s_j - x_j^2$.
Hence, for $1 \leq i \leq K$, we have the differential equations:
\begin{equation}
\frac{dx_i}{dt} = - i x_i -  i (s_i^2 - s_{i+1}^2)  + i \sum_{j + k = i} x_j
(s_k^2 - s_{k+1}^2).
\end{equation}
This family of equations captures the distribution up to the bound $K$;  the total
fraction of vertices in components of size larger than $K$ is captured by
$s_{K+1} = 1 - \sum_{i=1}^K x_i$.

To find the point where the phase transition occurs we consider
the evolution of the second moment of the component sizes, which we denote by $W$.  (Again, the dependence on $t$ is implicit.)  We define $W = \frac{1}{n}\sum_v c(v)$; 
this is the expected size of the component to which an arbitrary vertex belongs. 
We may also write $W = \sum_{i=1}^\infty i^2 n_i = \sum_{i=1}^\infty ix_i =
\sum_{i=1}^\infty s_i$.
It helps notationally to let $W^* = W -
\sum_{i=1}^K i x_i$; here $W^*$ corresponds to the contributions to
$W$ from vertices in components larger than the bound $K$.  Finally,
when components of size $j$ and $k$ are merged, the change in $W$ is equal to $(j+k)^2 - j^2 - k^2 = 2jk$. 
The evolution of $W$ is: 
\begin{eqnarray}\label{eqn:W_AE}
\frac{dW}{dt} & = & \sum_{j=1}^K \sum_{k=1}^K 2jk x_j (s_k^2 - s_{k+1}^2) + \sum_{j=1}^K 2j W^* x_j s_{K+1}  \nonumber \\ 
& & \mbox{} + \sum_{k=1}^K 2k W^* (s_k^2 - s_{k+1}^2) + 2 (W^*)^2 s_{K+1}.
\end{eqnarray}

The four terms can be explained as follows.
1) Both components have size $\le K$:  
the change in $W$ is $2jk$ multiplied by the 
respective probabilities that the first vertex has component size $j$ and the smaller of other two components has size $k$.
2) The first vertex has component size $j \leq K$;  the second is larger than $K$.  
(Note that the sum $\sum_{k=K+1}^\infty kx_k$ simplifies to $W^*$, which we have used to simplify the expression.)
3) The first vertex has component size greater than $K$ and the second does not. 
4) All three components have size greater than $K$. 

As $\sum_{j=1}^K j x_j = W - W^*$, we can simplify to obtain
\begin{equation}
\frac{dW}{dt} = 2 W \sum_{k=1}^K k (s_k^2 - s_{k+1}^2)
+ 2 W W^* s_{K+1},
\end{equation}
which can be used to show that this bounded size rule must
eventually reach a critical point where $W$ grows to infinity.  
Specifically, since $s_k \ge s_{k+1}$,
\begin{equation}
\frac{dW}{dt} \geq 2 W W^* s_{K+1}.
\end{equation}
Consider the first point where $s_{K+1} \geq \epsilon$ for some constant $\epsilon > 0$.  
(Since we keep adding edges, and $K$ is a constant, it
is straightforward to show that $s_{K+1}$ must eventually grow larger than 
a suitably small constant $\epsilon$.)
At this point $W^* \geq \epsilon W$ (since the at least $\epsilon$
fraction of $W$ from large components must contribute at least $\epsilon W$ of
$W$'s value), implying
$\frac{dW}{dt} \geq 2 \epsilon^2 W^2$, from which it follows that
$W$ goes to infinity at some finite time.  

It is tempting (but somewhat unrigorous) to consider the limiting version of
these eqns without the bound $K$: 
\begin{equation}\label{eq:dW_nobound}
\frac{dW}{dt} = 2W \sum_{k=1}^\infty s_k^2 > 2W.
\end{equation}
It is not immediately clear how to use Eq.~\ref{eq:dW_nobound} to
similarly demonstrate a critical point for the unbounded case.

Further details need to be dealt with to formalize the
accuracy of the differential equations; here we refer the reader to~\cite{SpencWorm}, which provides a full treatment for the case
where two independent edges are chosen for each step.  In particular,
a key issue is that the differential equations fail to take into
account redundant steps, where an edge joins two vertices that are
already in the same component.  The behavior for the $x_k$'s
is relatively straightforward under bounded-size
rules;  the probability that two vertices chosen at random
fall in the same component of size at most $K$ is $O(K^2/n)$, and the
asymptotic effect of such deviations does not affect convergence to
the differential equations.  The argument is more
challenging for bounding the effect on $W$, as $W$'s growth involves
components of size larger than $K$; however, by showing that the
fraction of vertices in components of size $k$ with high probability
eventually falls geometrically with $k$ (as detailed in~\cite{SpencWorm}), similar bounds can be shown to hold.  
 
One additional benefit of considering local schemes is that various generalizations are 
entirely transparent.  For example, the extension to $d$ choices of neighbors of the first vertex instead of two for a given integer $d$ yields
\begin{eqnarray}
  \frac{dx_i}{dt} = - i x_i -  i (s_i^d - s_{i+1}^d)  + i \sum_{j + k = i} x_j
(s_k^d - s_{k+1}^d);
\end{eqnarray}
\vspace{-0.2 in}
\begin{eqnarray}
\frac{dW}{dt} = 2 W \sum_{k=1}^K k (s_k^d - s_{k+1}^d) + 2 W W^* s_{K+1}^{d-1}.
\end{eqnarray}
Again, the limiting variation as $K$ goes to infinity has the simpler form
$\frac{dW}{dt} = 2 W \sum_{k=1}^\infty s_k^d.$
\smallskip


The second process we study (suggested
in~\cite{powderkeg09,PKpriv07}) is the triangle rule (TR): at each step choose
three distinct vertices uniformly at random, examine the triangle of
three possible edges connecting the pairs of vertices, and select the
edge that connects the two smallest components. 
Typical evolution of $C_1$ for this process
is shown in Fig.~\ref{fig:evol}. 
The bounded size rule variant is that if all components have size above
the bound $K$, we choose a random edge from the three;  if two components
have size above the bound $K$, we choose a random edge from the two
adjacent to the smallest component.
Using the same notation and analysis approach as for the AE rule, we can find differential equations for the bounded size variant; 
 \begin{eqnarray}\label{eq:xiTR}
 \frac{dx_i}{dt} & = & - 2ix_i^3 - 6ix_i^2s_{i+1} - 3ix_i^2(1-s_i) - 3ix_is_{i+1}^2  \nonumber \\
 & & \mbox{} - 6ix_is_{i+1}(1-s_i) + 6i \sum_{j + k = i; j < k } x_j x_k s_{k+1}
 \nonumber \\
 & & + \mbox{} ix_{i/2}^3  + 3i \sum_{j + k = i; j < k } x_j x_k^2
 + 3ix_{i/2}^2 s_{i/2+1}.
 \end{eqnarray} 

We briefly explain each term of Eq.~\ref{eq:xiTR}.  
1) All components have size $i$;  $2i$ vertices lost.  
2) Two components have size $i$, one has size greater than $i$. 
 3) Two components have size $i$, one has size less than $i$.  
4) One component has size $i$, and two have size larger than $i$.  
5) One component has size $i$, one has size less than $i$, 
one has size greater than $i$. 
6) All three components have different sizes, and the smallest two sum to $i$.
7) All three components have size $i/2$.  (This term only appears when $i$ is even.)
8) The two largest components have equal size, and the smallest two sum to $i$.
9) The two smallest components have equal size and sum to $i$.  (This term only appears when $i$ is even.)
Explaining, for example, the second term 
in more detail: we lose $2i$ vertices from $x_i$ when two components have size $i$ and one has size greater than $i$,
and the probability of this is $3x_i^2s_{i+1}$ when we take into account the orderings of the choices.  

 We again analyze how $W = \frac{1}{n}\sum_v c(v)$ changes:  
\begin{eqnarray}\label{eq:W_TR}
\frac{dW}{dt} & = & \sum_{j=1}^{K-1} \sum_{k=j+1}^K 12 jk x_j x_k s_{k+1} +  \sum_{j=1}^K \sum_{k=j+1}^K 6jk x_j x_k^2  \nonumber \\
             &   & \mbox{} + \sum_{j=1}^{K-1}  6j^2 x_j^2 s_{j+1} +  \sum_{j=1}^K  2j^2 x_j^3  \nonumber \\
&   & \mbox{}
+ W^* \sum_{j=1}^K  6jx_j s_{K+1}  + 2(W^*)^2 s_{K+1}.
\end{eqnarray}

The triangle setting lacks the pleasant form of the 
AE rule, but is suitable for calculation and fairly succinct. 
Here too, we can similarly create equations for merging the two largest components instead of the two smallest.

For both the adjacent edge (AE) and triangle rule (TR) models, we
solve the differential equations numerically using Euler's method in order to calculate,
roughly, the location of the phase transition.  
We 
discretized time with steps of size
$10^{-6}$.
More
sophisticated approaches using higher precision and error bounds could
yield more precise values, but the simple approach is sufficient for our current purposes. 
For the AE model, using a value of $K = 400$ led to an explosion in $W$ occurring between times $0.794$ and $0.795$; for $K = 600$, the explosion occurred slightly
later, between times $0.795$ and $0.796$.  For the TR model,
at $K = 400$ the explosion occurred between times $0.847$ and $0.848$,
and for $K = 600$ it occurred between times $0.848$ and $0.849$.
This closely matches the results from direct simulation of the graph evolution processes discussed next.

\begin{figure*}[tb]
  \begin{center}
  \includegraphics[width=\textwidth]{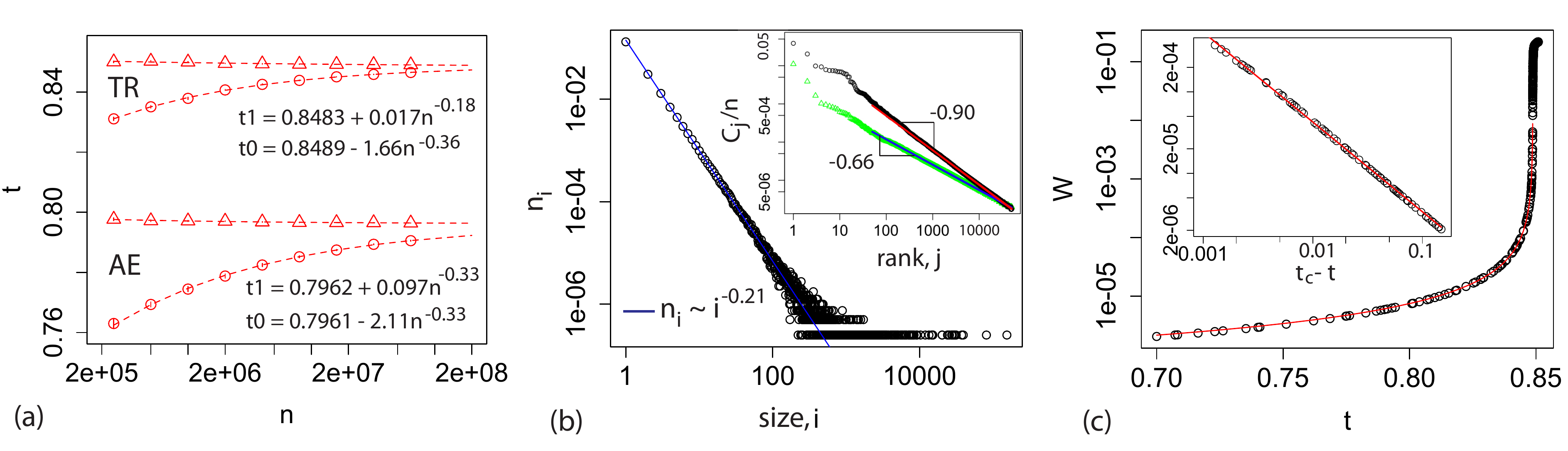}
  \end{center}
  \vspace{-0.2in}
  \caption{
(a) Measuring the lower and upper boundaries of $\Delta_n(1/2,0.2)/n$ for AE and $\Delta_n(1/2,0.4)/n$ for TR.
(b) Component density $n_i \sim i^{-2.1}$ shown for AE at $t_c$ (TR and PR are similar). More interesting is the rank-size component distribution (inset for ER and AE at $t_c$), showing the preponderance of large components for AE. Fitting for $50 < j < 50,000$ yields $C_j \sim j^{-\delta}$, with $\delta=-0.66$ for ER and $\delta=-0.90$ for AE (TR and PR are similar but more noisy). 
(c) $W$ versus $t$ for TR. 
Inset shows $W \sim (t_c-t)^{-\alpha}$ with red-line showing the best fit, attained with $\alpha=1.13$. The same red line is depicted in the main figure. 
}
\label{fig:crit}
\end{figure*}

We establish the explosive nature of the transition for both the
AE and TR models via numerical simulation of the underlying graph processes. 
We follow the approach introduced in~\cite{EPScience}, while here providing a more formal and detailed explanation of the procedure. 
Let $\Delta_n(\gamma,A)$ denote the number of edges required for 
$C_1$ to go from size $C_1 \le \lfloor n^\gamma\rfloor$ to size $C_1 \ge \lfloor An \rfloor$, for a system of $n$ vertices.  We wish to understand the asymptotic behavior, $\lim_{n \rightarrow \infty} \Delta_n(\gamma,A)$. If $\Delta_n(\gamma,A)$
increases linearly with $n$, then the time difference spanned by the
window, $\Delta_n(\gamma,A)/n$, approaches a limiting constant greater
than zero (the slope of $\Delta_n(\gamma,A)$ versus $n$).  If, in
contrast, $\Delta_n(\gamma,A) \propto n^\beta$ with $\beta < 1$ (\ie,
$\Delta_n(\gamma,A)$ is sublinear in $n$), then 
$\Delta_n(\gamma,A)/n \rightarrow 0$ as
$n\rightarrow \infty$.  In other words $C_1$ goes from size $n^\gamma$
to size $An$ in a time difference which approaches zero (shown for
AE and TR in Fig.~\ref{fig:crit}(a)).

As shown in the inset to Fig~\ref{fig:evol}, for the AE model we find that $\Delta_n(0.5,A) \sim n^{0.68}$ for all $A \in [0.1,0.3]$. For the TR model we find $\Delta_n(0.5,A) \sim n^{0.63}$ for all $A \in [0.1,0.4]$. The lower bound should decrease as we access larger $n$.  The upper bound estimates the largest value of $A$ for which the scaling is sublinear, denoted $A_c$.  Formally, $A_c = \sup_A \left[\lim_{n\rightarrow \infty} \Delta_n(\gamma,A)/n \rightarrow 0 \right]$, which is the size of the discontinuous jump in $C_1/n$ when viewed within this scaling window. 

We can bound the critical point for each process using the upper and lower boundaries of $\Delta_n(\gamma,A)$. Namely, we measure how $t_0$, the last time for which $C_1 \le n^\gamma$, and likewise how $t_1$, the first time for which $C_1 \ge An$, depend on $n$. As shown in Fig.~\ref{fig:crit}(a), we find that $t_0$ and $t_1$ approach essentially the same limiting value denoted $t_c$.  Neither of these local models is as effective in delaying the onset of the giant component as the original Product Rule (PR) studied in~\cite{EPScience} where the critical point $t_c \approx 0.888$.  For AE, $t_c \approx 0.796$, while for TR, $t_c\approx 0.848$.  Likewise neither model is as ``explosive" since $A_c\approx 0.6$ for PR, $A_c \approx 0.3$ for AE, and $A_c \approx 0.4$ for TR. 
Other well-known processes have now been shown to have 
discontinuous Achlioptas process counterparts~\cite{ziffPRL09,RadFortPRL09,ChoPRL09}. 

The discontinuous jump in the order parameter $C_1$ is characteristic of first order phase transitions.  Yet,  we observe critical scaling characteristic of second order transitions. Figure~\ref{fig:crit}(b) shows that $n_i$, the scaled number of components of size $i$, behaves as $n_i \sim i^{-\tau}$, with $\tau=2.1$ for both AE and TR (matching recently reported results for PR~\cite{ChoClusterAgg,RadFortPart2,Manna}). Figure~\ref{fig:crit}(c) shows how $W$ diverges at the critical point, behaving as $W \sim \left|t - t_c\right|^{-\alpha}$.  We also see similar behavior for the size of the second largest component, $C_2 \sim \left|t-t_c\right|^{-\mu}$.  Our numerical estimates are $\alpha = \mu \approx 1.13$ for AE and TR, while $\alpha=\mu\approx 1.17$ for PR. 
Note, we recover the standard \ER \ (ER) exponents ($\tau=5/2$ and $\alpha=\mu=1$) in our simulations of ER. Hybrid phase transitions 
have been previously observed 
for spin glasses~\cite{GrossMezard,thiru}, constraint satisfaction
problems (K-SAT)~\cite{MonasEtal99}, models of jamming in granular materials 
(see~\cite{liunagel,
knights06,SchwarzLiuChayes06} and references therein), and $k$-core percolation~\cite{kcore}.


In summary we have introduced local models of graph evolution with choice that can be described by mathematical evolution equations and which exhibit  
discontinuous percolation transitions with critical scaling behaviors. Discontinuous percolation transitions are not yet fully understood. Local processes appear much simpler to describe mathematically 
and thus offer the potential for a system with a discontinuous percolation transition that is easier to analyze.

We thank Microsoft Research New England where this research was initiated.  R.D. was supported in part by National Academies Keck Futures Initiative grant CS05 and M.M. by NSF grants CCF-0634923 and CCF-0915922.

\end{document}